\documentclass[aps,twocolumn,prl,superscriptaddress]{revtex4-1}
\usepackage{longtable}
\usepackage{morefloats}
\usepackage[dvips]{graphicx}
\usepackage{color}
\usepackage{epsfig,graphicx,amsfonts,amsbsy}
\usepackage{amsmath,amsfonts,amsthm,amssymb}
\usepackage{appendix}
\usepackage{makeidx}
\usepackage{url}
\usepackage[bookmarksnumbered,pdfpagelabels=true,plainpages=false,colorlinks=true,linkcolor=blue,citecolor=blue,urlcolor=blue]{hyperref}

\begin{document}
\title{Flat Bands, Indirect Gaps, and Unconventional Spin-Wave Behavior Induced by a Periodic Dzyaloshinskii-Moriya Interaction}
\author{R. A. Gallardo}
\affiliation{Departamento de F\'{i}sica, Universidad T\'{e}cnica Federico Santa Mar\'{i}a, Avenida Espa\~{n}a 1680, Valpara\'{i}so, Chile}
\affiliation{Center for the Development of Nanoscience and Nanotechnology (CEDENNA), 917-0124 Santiago, Chile}
\author{D. Cort\'{e}s-Ortu\~{n}o}
\affiliation{Faculty of Engineering and the Environment, University of Southampton, Southampton SO17 1BJ, United Kingdom}
\author{T. Schneider}
\affiliation{Helmholtz-Zentrum Dresden -- Rossendorf, Institut of Ion Beam Physics and Materials Research, Bautzner Landstr. 400, 01328 Dresden, Germany}
\affiliation{Department of Physics, Technische Universit\"at Chemnitz, Reichenhainer Str. 70, 09126 Chemnitz, Germany}
\author{A. Rold\'an-Molina}
\affiliation{Universidad de Ays\'en, Calle Obispo Vielmo 62, Coyhaique, Chile}
\author{Fusheng Ma}
\affiliation{Magnetoelectronic Lab, School of Physics and Technology, Nanjing Normal University, Nanjing 210023, China}
\author{R. E. Troncoso}
\affiliation{Departamento de F\'{i}sica, Universidad T\'{e}cnica Federico Santa Mar\'{i}a, Avenida Espa\~{n}a 1680, Valpara\'{i}so, Chile}
\affiliation{Center for Quantum Spintronics, Department of Physics, Norwegian University of Science and Technology, NO-7491 Trondheim, Norway}
\author{K. Lenz}
\affiliation{Helmholtz-Zentrum Dresden -- Rossendorf, Institut of Ion Beam Physics and Materials Research, Bautzner Landstr. 400, 01328 Dresden, Germany}
\author{H. Fangohr}
\affiliation{Faculty of Engineering and the Environment, University of Southampton, Southampton SO17 1BJ, United Kingdom}
\affiliation{European XFEL GmbH, Holzkoppel 4, 22869 Schenefeld, Germany}
\author{J. Lindner}
\affiliation{Helmholtz-Zentrum Dresden -- Rossendorf, Institut of Ion Beam Physics and Materials Research, Bautzner Landstr. 400, 01328 Dresden, Germany}
\author{P. Landeros}
\affiliation{Departamento de F\'{i}sica, Universidad T\'{e}cnica Federico Santa Mar\'{i}a, Avenida Espa\~{n}a 1680, Valpara\'{i}so, Chile}
\affiliation{Center for the Development of Nanoscience and Nanotechnology (CEDENNA), 917-0124 Santiago, Chile}

\date{\today }

\begin{abstract}
Periodically patterned metamaterials are known for exhibiting wave properties similar to the ones observed in electronic band structures in crystal lattices.   
In particular, periodic ferromagnetic materials are characterized by the presence of bands and bandgaps  in their spin-wave spectrum at tunable GHz frequencies.   
Recently, the fabrication of magnets hosting Dzyaloshinskii-Moriya interactions has been pursued with high interest since properties such as the stabilization of chiral spin textures and nonreciprocal spin-wave propagation emerge from this antisymmetric exchange coupling. 
In this context, to further engineer the magnon band structure, we propose  the implementation of magnonic crystals with periodic Dzyaloshinskii-Moriya interactions, which can be obtained, for instance, via patterning of periodic arrays of heavy-metals wires on top of an ultrathin magnetic film. 
We demonstrate through theoretical calculations and micromagnetic simulations that such systems show an unusual evolution of the standing spin waves around the gaps in areas of the film that are in contact with the heavy-metal wires. 
We also predict the emergence of indirect  gaps and flat bands and, effects that depend on the strength of the Dzyaloshinskii-Moriya interaction.
This study opens new routes towards engineered metamaterials for spin-wave-based devices.

\end{abstract}
\maketitle


Exotic wave phenomena observed in artificial periodic structures have made the study of metamaterials an active research field that spans different areas of condensed matter physics such as photonics, phononics, plasmonics, and magnonics~\cite{Zheludev12,Hess12,Maldovan13,Yu13,Chumak15,Lodahl17}. 
Photonic crystals lacking space-inversion and time-reversal symmetries were shown to exhibit nonreciprocal dispersions with interesting consequences, even in one-dimensional systems \cite{Eritsyan00,Koerdt03,Gevorgyan02,Bita05,Lodahl17}. 
Magnetic metamaterials, referred to as magnonic crystals (MCs), are obtained by periodically modulating the properties of the host materials, which in turn enables the manipulation of spin waves (SWs)~\cite{Vasseur96,Nikitov01,Kruglyak06,Krawczyk08,Lee09,Neusser09,Kruglyak10,Serga10,Gubbiotti10,Kim10,Wang10,Khitun10,Lenk11,Demokritov12,Krawczyk14,Tacchi17}.
Such waves exhibit GHz  frequencies with sub-micrometric  wavelengths, which are significantly shorter than those of electromagnetic waves of the same frequency observed in photonic crystals~\cite{Joannopoulos2011}. 
For practical reasons, this important feature encourages the miniaturization of magnonic devices~\cite{Tacchi15}.  
Brillouin light scattering experiments, together with theoretical and numerical studies, have confirmed the presence of bandgaps in MCs~\cite{Demokritov12,Krawczyk14,Tacchi17}.
The filtering of SWs at specific frequency ranges allows to perform logic operations, signal processing or the realization of magnon transistors, thus offering novel prospects for information and communication technologies~\cite{Khitun05,Jamali13,Kim09,Chumak14}.

The theoretical prediction~\cite{Dzyaloshinskii58,Moriya60,Fert80} and subsequent experimental probe of magnets with Dzyaloshinskii-Moriya interaction (DMI) enabled the observation of chiral spin textures such as helices, spirals and skyrmions in these materials~\cite{Roszler06,Muhlbauer09,Nagaosa13,Wiesendanger16,Fert17}. 
Moreover, multiple studies have established that magnons are nonreciprocal in such systems~\cite{Udvardi09,Zakeri10,Costa10,Cortes13,DiPRL15,Cho15,Nembach15,Belmeguenai15,Tacchi17b}, especially in thin ferromagnetic (FM) films next to a strong spin-orbit-coupling material, such as a heavy metal (HM), where the broken space-inversion symmetry leads to an interfacial DMI~\cite{Fert17,Wiesendanger16}. 
This fact, together with the broken time-reversal symmetry in  ferromagnets, encourages to anticipate that a rich variety of physical phenomena arises from the combination of magnonics and interfacial DMI.  
Indeed, nonreciprocal devices have the ability to transfer energy unidirectionally \cite{An13} and can be used as microwave isolators and circulators \cite{Verba13,Krawczyk14}. 
Unidirectional propagation in a narrow frequency band was shown in bi-component MCs in contact with a heavy-metal layer~\cite{Ma14}. 
The DMI also shifts the frequencies of the spin-wave modes and increases the intensity of the absorption peaks of the calculated ferromagnetic resonance (FMR) spectrum of one-dimensional MCs~\cite{Mruczkiewicz16}. 
Additional  modes in the FMR spectrum of FM/HM ellipses have been recently predicted using computer simulations, where the confinement creates a fixed nodal structure, generating an unusual SW behavior~\cite{Zingsem17}.  
The nature of the magnon bands was theoretically discussed in an array of magnetic nanoislands on top of a HM layer, a system that admits for topologically nontrivial bands~\cite{Iacocca17}.
Spin waves can be also amplified at the boundary between two regions with different DMI~\cite{Lee17}, while a similar system was chosen to study domain wall motion~\cite{Hong17}. 
Although these works highlight key aspects of the interplay between spin waves and DMI,  exceptional features resulting from a \emph{periodic} Dzyaloshinskii-Moriya coupling have been overlooked so far.

In this letter, we study the role of a periodic interfacial DMI on the magnon band structure of ultrathin films.  
We theoretically describe these systems with the plane-wave method~\cite{Krawczyk08} and using the MuMax3~\cite{Vansteenkiste2014} and OOMMF~\cite{Donahue99} micromagnetic codes~\cite{SUPPMAT}. 
We predict that a periodic modulation of the Dzyaloshinskii-Moriya strength induces three main effects: 
(i) Tunable indirect bandgaps, 
(ii) Low-frequency flat magnonic bands, 
and (iii) An unconventional temporal evolution of the standing spin waves in the areas with active DMI underneath the heavy-metal wires.  
While the indirect bandgap may be expected due to the nonreciprocity, the time evolution of such modes is an uncommon phenomenon.
This effect implies that SWs have the ability to propagate in areas covered with HM wires and, at the same time, retain their standing wave character in the  uncovered regions. 
The existence and further control of such a hybrid kind of wave, with both standing and propagating features, in turn could push the physics of spin waves further towards a new generation of devices for information technologies.

\begin{figure}[t]
\includegraphics[width=0.7\columnwidth]{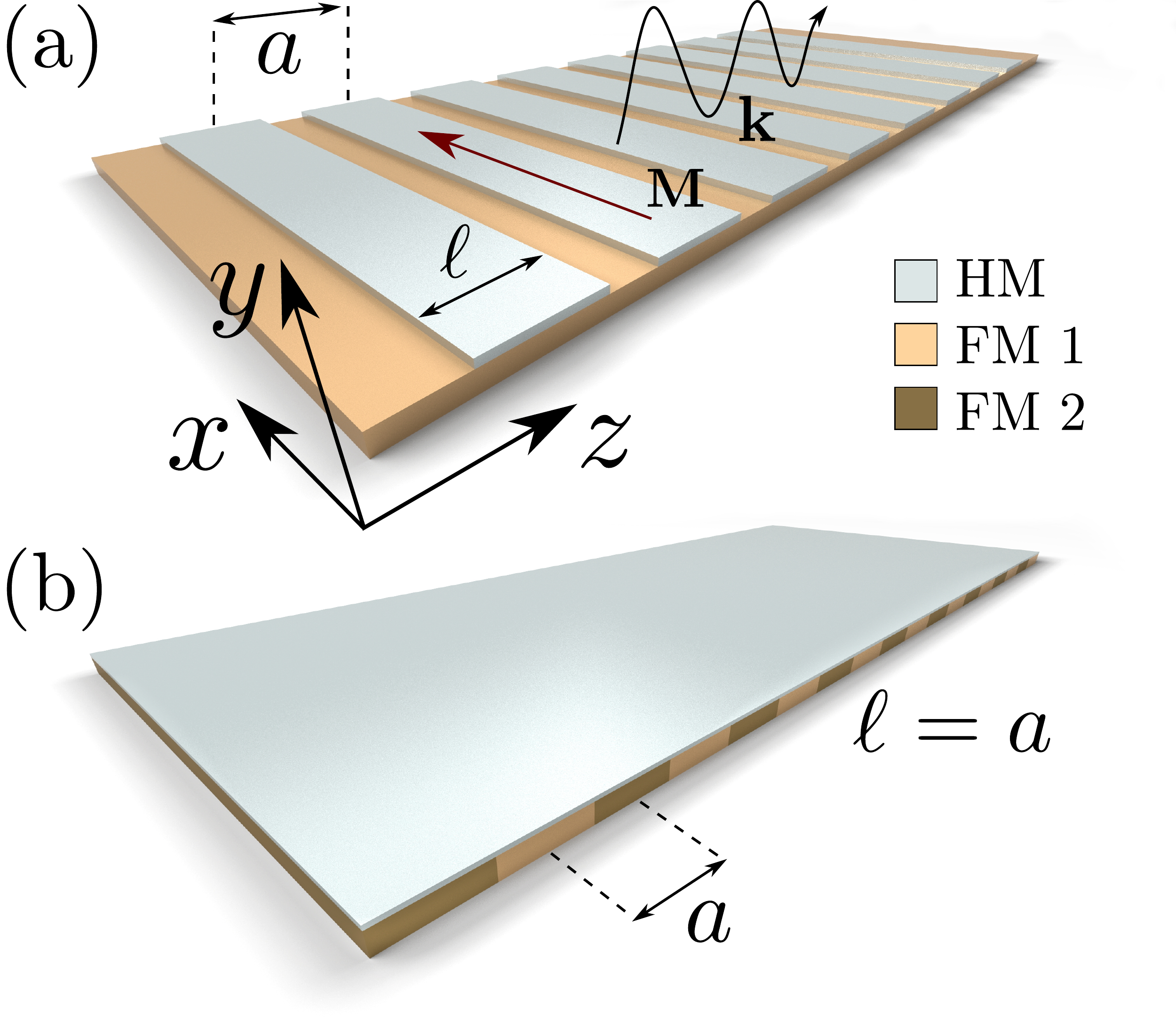}
\caption{Illustration of two possible realizations of a chiral magnonic crystal where spin waves propagate along $z$ and the equilibrium magnetization points along $x$. 
(a) A ferromagnetic ultrathin film covered with a periodic array of heavy-metal wires such as Pt.
(b) A bicomponent magnonic crystal composed of ferromagnets 1 and 2  is coupled to a heavy-metal layer.
Both systems induce a periodic Dzyaloshinskii-Moriya interaction at the interface. 
}
\label{Fig1}
\end{figure}

The envisioned chiral magnonic crystals with a periodic DMI are illustrated in Fig.~\ref{Fig1}, where a thin FM film is covered with an array of HM wires, and a bi-component MC is coupled to a HM layer. 
Here, $a$ is the periodicity and $\ell$ the width of the  wires. 
We describe the SW dynamics using the Landau-Lifshitz equation taking into account exchange, dipolar, Zeeman and Dzyaloshinskii-Moriya interactions.  
According to Bloch's theorem and assuming SWs propagating along $z$, the dynamic magnetization (together with the saturation magnetization $M_{\rm s}$, exchange length $\lambda$, and Dzyaloshinskii-Moriya constant $D$) is expanded into a Fourier series~\cite{SUPPMAT}.
Since the DMI depends on the position along $z$, the derivation of the corresponding field in the equation of motion requires a careful analysis, which can be found in Sec.~S1~\cite{SUPPMAT}.
This periodic DMI also leads to twisted boundary conditions  \cite{Rohart13,Garcia14} for the magnetization at the interfaces of zones with and without DMI (see Sec.~S2~\cite{SUPPMAT}).
Additionally, we solve the spin-wave modes through micromagnetic simulations (see Sec. S3~\cite{SUPPMAT}). 

\begin{figure*}[ht]
\includegraphics[width=0.75\textwidth]{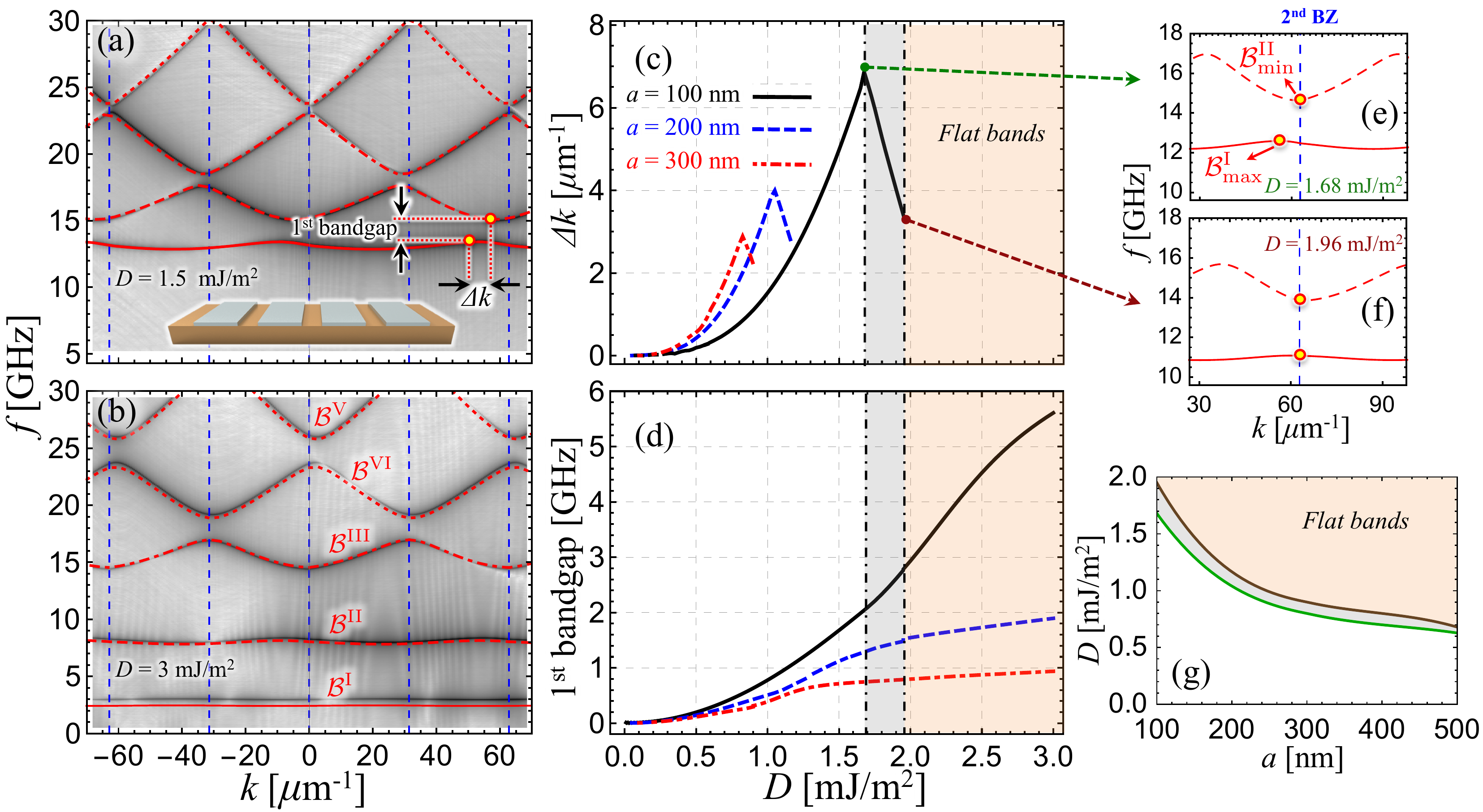} 
\caption{(a,b) Magnon band structure for a Ni$_{80}$Fe$_{20}$ film covered with a HM-wire array, where $a=100$~nm, $D=1.5$ and $3$~mJ/m$^2$. 
The lines correspond to the theory and the color code to the MuMax3 results, where darker (lighter) color represents an intensity maximum (minimum).
(c,d) Depicts the wavevector shift $\Delta k$ and the 1$^{\rm{st}}$ bandgap as a function of the Dzyaloshinskii-Moriya strength $D$, where the solid, dashed, and dot-dashed lines represent the periodicities $a=100$, $200$ and $300$~nm, respectively.
The shaded areas correspond to three ranges of $D$-values  where different behaviors are predicted.  
(e,f) The first two spin-wave branches at the critical values $D=1.68$ and $1.96$~mJ/m$^2$, for $a=100$~nm. 
In (g) the three regions of $D$-values are shown as a function of the periodicity of the HM-wire array. 
}
\label{Fig2}
\end{figure*}

For the system shown in Fig.~\ref{Fig1}(a), we use a  permalloy (Ni$_{80}$Fe$_{20}$) film with  $M_{\rm s} = 658\,\text{kA/m}$,  $\lambda=6.39\,\text{nm}$, gyromagnetic ratio $\gamma=175.87\,\text{GHz/T}$, and thickness $d=3$~nm.
Spin-wave dispersion curves are shown in Fig.~\ref{Fig2}(a-b) for $a=100\,\text{nm}$,  $\ell=a/2$, and  $D=1.5$ and $3\,\text{mJ/m}^2$, which are close to reported values \cite{Belmeguenai15,Tacchi17b}.
A bias field $\mu_0 H= 250\,\text{mT}$ is applied along $x$ and the SWs propagate along $z$.
The intensity plots in the background are the results from MuMax3 simulations whilst the lines correspond to the theory, where we have labeled the frequency branches as $\mathcal{B}^{\text{I}}$, $\mathcal{B}^{\text{II}}$, etc.  
In the case $D=0$ (see Fig. S3(b) in \cite{SUPPMAT}), the SW dispersion is reciprocal with the lowest frequency $f\approx 14.6$~GHz at $k=0$. 
By increasing $D$, the SW dispersion becomes nonreciprocal, magnonic gaps open due to the periodic DMI, and the bottom of the dispersion shifts down in frequency and towards higher wave-vectors for $k>0$ \footnote{Note that the $k=0$ state does not change its frequency with $D$, but the group velocity does.}. 
Interestingly, we observe indirect bandgaps, which emerge out of the edges of the Brillouin zones (BZs), which are denoted by vertical dashed lines in Fig.~\ref{Fig2}.
Indirect bandgaps originated by the inhomogeneous distribution of the surface SW intensity across the thickness, have been reported on thick metallized MCs with periodicities of hundreds of micrometers \cite{Mruczkiewicz13,Mruczkiewicz14,Bessonov15}. 
Then, the associated wavelengths are three orders of magnitude smaller in our system, which imposes a significant advantage for the miniaturization of magnonic devices.
Such gaps are due to nonreciprocity, since different wavelengths of two counter-propagating waves shift the minimum and maximum of a given branch out of the borders of the BZs.
For $D>0$, the maximum of the 1$^{\rm{st}}$ branch ($\mathcal{B}^{\rm I}_{\rm max}$) and the minimum of the 2$^{\rm{nd}}$ branch ($\mathcal{B}^{\rm II}_{\rm min}$) move to the right, but in a different way, which gives rise to the indirect bandgaps. 
The same behavior was obtained using the MuMax3 and OOMMF codes (see Fig.~\ref{Fig2} and Fig. S3 in \cite{SUPPMAT}).

The dependence of the indirect bandgaps with respect to the DMI strength is shown in Fig.~\ref{Fig2}(c), where we choose $\ell=a/2$ with $a=100$, $200$ and $300\,\text{nm}$. 
Further, in Fig.~\ref{Fig2}(d) we plot the 1$^{\rm{st}}$ bandgap, 
which clearly increases with $D$. 
The wavevector separation $\Delta k$ between the $k$-values at the minimum of $\mathcal{B}^{\rm II}$ and maximum of $\mathcal{B}^{\rm I}$, evidence a nonmonotonic response with $D$.
Such behavior motivates us to define three ranges of $D$: weak (white region), intermediate (gray region), and strong (flat  bands region). 
The boundaries between these values depend on periodicity and have been highlighted just for  $a=100$ nm in Figs.~\ref{Fig2}(c,d).
For weak $D$, $\Delta k$ and the gap increase with it, up to $D\approx 1.68$~mJ/m$^2$, where $\mathcal{B}^{\rm II}_{\rm min}$ reaches the second BZ and remains pinned at its corresponding $k$ value [see Fig.~\ref{Fig2}(e)].  
In the intermediate region ($1.68 <D<1.96$~mJ/m$^2$),  the minimum of the second branch still remains pinned at the 2$^{\rm{nd}}$~Brillouin zone while  $\mathcal{B}^{\rm I}_{\rm max}$ moves towards such $k$-point, which explains the evident reduction of $\Delta k$ in the gray region of Fig.~\ref{Fig2}(c).  
As the maximum of the first branch reaches the 2$^{\rm{nd}}$~Brillouin zone [see Fig.~\ref{Fig2}(f)], the branch  becomes flat and defines the strong $D$-range. 
If $D$ further increases, the upper-frequency branches also become flat \footnote{That is the reason why $\Delta k$ is not shown for strong $D$ in Fig.~\ref{Fig2}(c), where the branch $\mathcal{B}^{\rm I}$ basically does not have a maximum and hence, $\Delta k$ is not well defined.}.
For other periodicities, the boundaries between such $D$-values are displayed in Fig.~\ref{Fig2}(g), where the reduction of these boundaries with increasing periodicity becomes evident.
The latter implies that larger periodicities favor the formation of flat bands, as a consequence of a reduction of the coupling between SWs underneath different HM wires. On the other hand, shorter periodicities induce larger values of $\Delta k$.
Magnon flat bands have been observed in crystalline spin systems  \cite{Kageyama00,Matan06,Chisnell15,dAmbrumenil15,Janoschek10,Kugler15}, magnonic crystals \cite{Gubbiotti05,Zhang12,Pan13}, and magnonic superlattices \cite{Di14,Gallardo18}, where only Refs. \cite{Chisnell15,dAmbrumenil15,Janoschek10,Kugler15,Pan13} show flat bands as the low-lying excitations. 
However, these systems having flat bands do not show indirect gaps nor the unusual behavior of the spin waves reported in our manuscript. 
Since the flat modes exhibit a significant intensity in a broad $k$-range, they should be observed by Brillouin light scattering as well as by ferromagnetic resonance. 
Although it has been predicted that for uniformly magnetized infinite thin films the DMI does not modify the FMR mode~\cite{Cortes13}, under the influence of a periodic DMI the FMR modes are clearly affected \cite{Mruczkiewicz16}.
The emergence of the flat magnonic bands is enhanced as the periodicity increases (not shown), and the SW degeneracy may have interesting consequences, as the Bose-Einstein condensation predicted in  bosonic systems hosting flat bands \cite{Huber10}, and frustrated quantum magnets \cite{Gardner10,Zapf14}.
Additionally, topological properties, such as chiral edge states and non-trivial magnon conductivities, could be observed in higher-dimensional periodic DMI structures, where the Berry phase becomes relevant  \cite{Shindou13}.

\begin{figure*}[ht]
\includegraphics[width=0.8\textwidth]{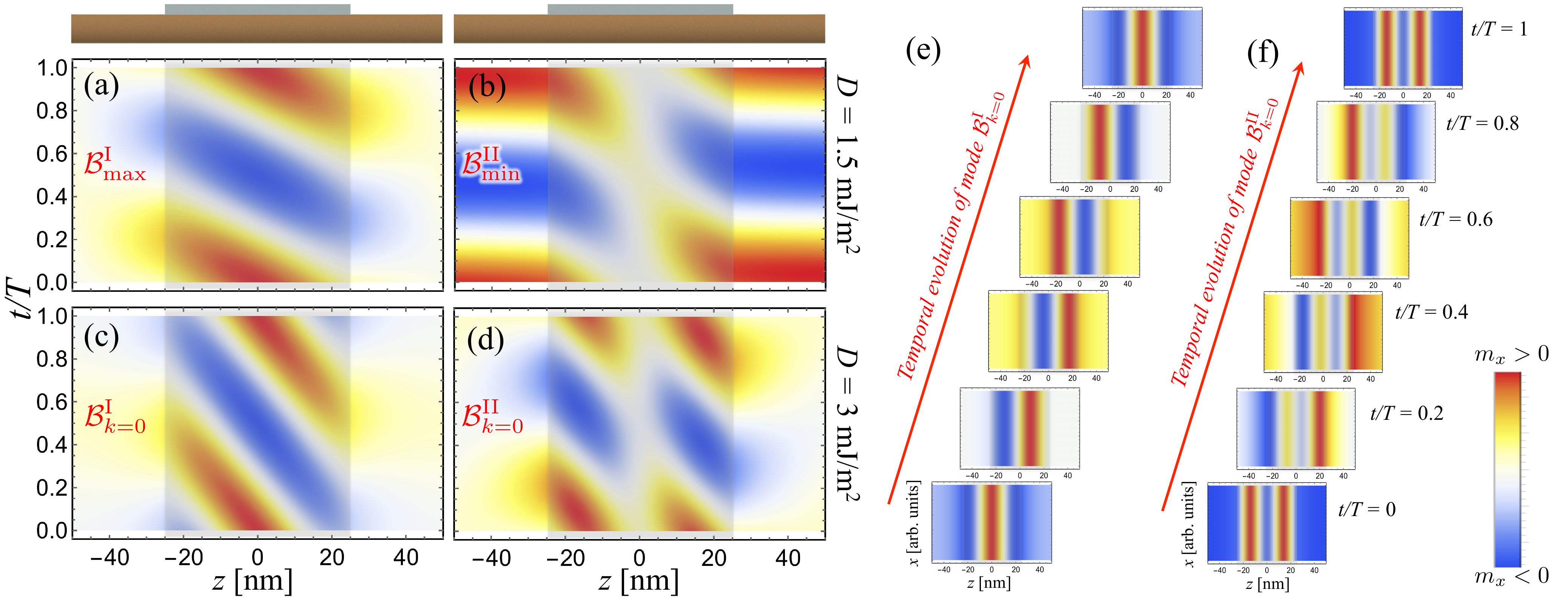}
\caption{Temporal and spatial evolution of the spin-wave modes, where $T$ is the period of the oscillation. (a,b) Profiles of the modes $\mathcal{B}^{\rm I}_{\rm max}$ and $\mathcal{B}^{\rm II}_{\rm min}$,  for $D=1.5$ mJ/m$^2$. 
(c,d) Mode profiles for the first and second branches ($\mathcal{B}^{\rm I}$ and $\mathcal{B}^{\rm II}$) evaluated at $k=0$ for $D=3$ mJ/m$^2$.  (e,f) Snapshots of the temporal evolution of the modes $\mathcal{B}^{\rm I}_{k=0}$ and $\mathcal{B}^{\rm II}_{k=0}$.
}
\label{Fig3}
\end{figure*}

It has been previously established that the periodicity of the internal field in a MC, which gives rise to counterpropagating ``Bragg-reflected" waves, induces frequency bandgaps~\cite{Zivieri12}, where the superposition of two opposing waves creates standing SWs. 
However, in a chiral MC two counterpropagating SWs excited at the same frequency have different wavelengths, and therefore neither standing SWs nor bandgaps appear at the borders of the Brillouin zones, as shown in Fig.~\ref{Fig2}. 
Accordingly, the time evolution of the flat SW modes and the ones around the gaps are highly nontrivial and endemic of the periodic DMI.  
We emphasize this unique feature in Fig.~\ref{Fig3}, by calculating $m_x$ for different modes. 
Figs.~\ref{Fig3}(a,b) correspond to modes $\mathcal{B}^{\rm I}_{\rm max}$ and $\mathcal{B}^{\rm II}_{\rm min}$, for $D=1.5\,\text{mJ/m}^2$. 
These modes are neither symmetric nor antisymmetric and correspond to standing waves with local phase shifts. 
The dephasing occurs underneath the HM wires, and yields the nontrivial evolution of the magnetization displayed in Fig.~\ref{Fig3}. 
Below the HM wires (shaded areas) the nonreciprocal nature of the coupling creates modes with nonzero phase velocities, i.e., waves that propagate inside the regions with active DMI. 
In contrast, the usual standing character is observed outside these regions, hence the group velocities of the corresponding modes are zero. 
This unique behavior can be understood by analyzing the dynamic magnetization across the regions with and without DMI, which is governed by the boundary conditions~\cite{Rohart13,Garcia14} discussed in Sec. S2, where it is clear that DMI induces canted orbits (see Fig. S2~\cite{SUPPMAT}), explaining the particular spatio-temporal evolution of the SWs.
This exotic evolution of the SWs has not been reported before in any kind of MC. Only FMR simulations in a confined structure with DMI~\cite{Zingsem17} 
have shown a similar behavior, but limited to the $k=0$ case. 

In Figs.~\ref{Fig3}(c,d), the SW profiles for the $1^{\rm{st}}$ and $2^{\rm{nd}}$ branches are depicted for $D=3\,\text{mJ/m}^2$. 
Although these modes are evaluated at $k=0$, it can be shown that their profiles are independent of $k$. 
We also observe nonzero phase velocities in the zones with active Dzyaloshinskii-Moriya coupling, while the wave amplitudes are reduced in the regions without DMI, evidencing an evanescent character of the SWs.
The time evolution of both modes is shown in Figs.~\ref{Fig3}(e,f), where the propagating character in the region with active DMI can be seen.
Note that the slopes in the spatio-temporal plots [Fig.~\ref{Fig3}(a-d)] increases with $D$ and are zero for $D=0$, which can be understood assuming a weak DMI strength. 
Since standing SWs arises from the superposition of two opposing waves, in the region where the DMI is nonzero, the right wavevector has the form $k_R=k+k'$, while the left one is $k_L=-(k-k')$. 
Thus, the wavevector shift is $k'\propto D$~\cite{Zingsem17}. 
Due to the constant slope in the $t/T$ vs. $z$ plots, the phase velocity of the SWs underneath the HMs is $v_p=\Delta z/t$. 
Then, taking into account that $v_p=\omega/k'$, one can show that $\frac{t/T}{\Delta z}\propto D$, where $T$ is the period.
Then, the slopes become larger as $D$ increases. 

The modulation of the magnon band structure can be reached through the change of the width of the HM wires, as shown in Fig.~S3~\cite{SUPPMAT}, where an increment in the width decreases the frequency of the first spin-wave modes, until they become flat.
Alternatively, this can be obtained by using a bicomponent magnonic crystal in contact with a continuous heavy-metal layer (see Fig.~\ref{Fig1}b), where it is also possible to observe the emergence of flat bands, as shown in Fig.~S4~\cite{SUPPMAT}.
For both kind of chiral MCs, results are predicted from the theory and substantiated with OOMMF and MuMax3 simulations.


We have predicted remarkable effects  driven by a periodic Dzyaloshinskii-Moriya coupling, where we highlight the formation of indirect bandgaps, the nontrivial temporal evolution of standing spin waves around, and the formation of low-frequency flat bands. 
This has been achieved using  micromagnetic simulations and  a theoretical approach based on the plane-wave method, which describes the dynamic properties of magnonic systems with incorporated periodic interfacial Dzyaloshinskii-Moriya intyeraction.
The occurrence of indirect gaps is a consequence of the nonreciprocal nature of the SWs caused by DMI, and occurs for $D$ smaller than a critical value that depends on the periodicity of the wire array. 
For stronger $D$, we found flat magnonic bands that should be observable by Brillouin light scattering as well as by FMR measurements, and may be a platform for Bose-Einstein condensation of magnons \cite{Demokritov06}. 
Nonreciprocity also produces a unique temporal behavior of the SWs, with standing waves exhibiting finite phase velocities in the regions where the DMI is nonzero. 
We have demonstrated that chiral magnonic crystals host interesting physical properties, encouraging future experimental studies to prove and evidence these phenomena.

This work was supported by FONDECYT 11170736, 1161403, 3170647, CONICYT PAI/ACADEMIA 79140033, CONICYT Becas Chile 72140061, and Centers of excellence with Basal/CONICYT financing, grant FB0807, CEDENNA. 
D. C.-O. and H. F. acknowledge the EPSRC Programme grant on Skyrmionics (EP/N032128/1) and EPSRC’s Doctoral Training Centre in Complex System Simulation~(EP/G03690X/1). 
T.S. acknowledges funding from the InProTUC scholarship.


\end{document}